\documentstyle[aps,epsfig,amsmath,amssymb,array]{revtex}

\begin{document}

\title{Transition from oscillatory to excitable regime in a system forced at
three times its natural frequency}
\author{Rafael Gallego$^1$, Daniel Walgraef$^{\,2,1}$, Maxi San Miguel$^1$ and
Ra\'ul Toral$^1$}
\address{1.- Instituto Mediterr\'aneo de Estudios Avanzados, IMEDEA (CSIC-UIB),
Campus UIB, 07071-Palma de Mallorca, Spain\\
2.- Centre for Non-Linear Phenomena and Complex Systems, Universit\'e Libre de
Bruxelles, Campus Plaine, Blv. du Triomphe B.P 231, B-1050 Brussels, Belgium}
\maketitle

\begin{abstract}

The effect of a temporal modulation at three times the critical
frequency on a Hopf bifurcation is studied in the framework of
amplitude equations. We consider a complex Ginzburg-Landau
equation with an extra quadratic term, resulting from the strong
coupling between the external field and the unstable modes. We
show that, by increasing the intensity of the forcing, one passes
from an oscillatory regime to an excitable one with three
equivalent frequency locked states. In the oscillatory regime,
topological defects are one-armed phase spirals, while in the
excitable regime they correspond to three-armed excitable
amplitude spirals. Analytical results show that the transition
between these two regimes occurs at a critical value of the
forcing intensity. The transition between phase and amplitude
spirals is confirmed by numerical analysis and it might be observed
in periodically forced reaction-diffusion systems.

\end{abstract}

\vskip 0.4cm

\section{Introduction}

In many cases, the nucleation of spatio-temporal patterns is
associated with continuous symmetry breaking, and these patterns
are thus very sensitive even to small perturbations or external
fields. Perturbations may be induced by imperfections of the
system itself (e.g.  impurities), of the geometrical set-up (e.g.
the boundary conditions),  of the control parameters, etc. In
addition, external fields may  induce spatial or temporal
modulations of the control or bifurcation parameters. In fact,
spatially or temporally modulated systems are very common in
nature, and the effect of external fields on these systems has
been studied for a long time. As a way of example, the forcing of
a large variety of nonlinear oscillators, from the pendulum to Van
der Pol or Duffing oscillators, has led to detailed studies of the
different temporal behaviors that can be obtained. It has been
shown that  resonant couplings between the forcing and the
oscillatory modes may lead to  several types of complex dynamical
behaviors, including quasi-periodicity,  frequency lockings,
devil's staircases, chaos and intermittency
\cite{nonlinearosc,haobailin}.

In equilibrium systems, the importance of spatial modulations has been known
for  a long time. For example, in the case of spatial modulations occurring in
equilibrium crystals, such as spin or charge density waves, the constraint
imposed by the periodic structure of the host lattice leads to the now commonly
known commensurate-incommensurate phase transitions. The transition from the
commensurate phase, where the wavelength of the modulated structure is a
multiple of the lattice constant, to the incommensurate one occurs via the
nucleation of domain walls separating domains which are commensurate with the
host lattice \cite{bak80,bulaev78}.

In nonequilibrium systems, the systematic study of the influence
of external fields on pattern forming instabilities is more
recent. It has been first devoted to instabilities leading to
spatial patterns. For example, the Lowe-Gollub experiment
\cite{lowe85} showed that, in the case of the electrohydrodynamic
instability of liquid crystals, a spatial modulation of the
bifurcation parameter may induce dis-commensurations,
incommensurate wavelengths and domain walls. The similarities with
analogous equilibrium phenomena rely on the fact that, close to
this instability, the asymptotic  dynamics is described by the
minimization of a potential \cite{lubensky,coullet86}.

In the case of self-oscillatory systems, however, original effects
occur as a consequence of the nonrelaxational character of the
dynamics. In particular, for wave bifurcations, unstable standing
waves or two-dimensional wave patterns may be stabilized by pure
spatial or temporal modulations of suitable wavelengths or
frequencies \cite{riecke88,walgraef88,rehberg88,coulletwal89}. The
case of pure temporal modulations in oscillating extended systems
has been considered theoretically \cite{kjartan,Elphick} and in
experiments on chemical systems forced by periodic illumination
\cite{Swinney97,Swinney99,Swinney00,Swinney01}. The study by
Coullet and Emilsson of forced Hopf bifurcations is based on
amplitude equations of the scalar Ginzburg-Landau type. They
considered periodic temporal modulations of frequency $\omega_e =
(n/m)(\omega_0 - \nu )$, where $(n/m)$ is an irreducible integer
fraction, $\omega_0$ is the critical frequency of the Hopf
bifurcation, and $\nu $ is a small frequency shift. Such forcings
break the continuous time translation down to discrete time
translations, and the corresponding amplitude equations become:

\begin{equation}\label{gleq}
\partial_t A= (\mu+i\nu) A + (1+i\alpha ) \nabla^2 A - (1+i\beta )A\vert A\vert^2
+ \gamma_n\bar A^{n-1},
\end{equation}
($\bar A$ stands for the complex conjugate of $A$). In the non
forced case ($\gamma_n=0$) and for zero frequency shift ($\nu=0$)
this dynamical system is of the relaxational type for
$\alpha=\beta=0$. In more general situations the system follows a
nonrelaxational dynamics\cite{non-pot}. If the forcing
intensity $\gamma_n$ is sufficiently strong, this dynamics admits
asymptotically stable uniform steady states, corresponding to
frequency locked solutions. There are $n$ different frequency
locked solutions, which only differ by a phase shift of $2\pi /n$.
These solutions are always stable in the large forcing limit for
$n=1,2,3,4$, the so-called strongly resonant cases. In this
regime, the dynamics resembles some sort of excitability. The
locked solutions may undergo various types of instabilities
\cite{kjartan}. One of them is of phase type and occurs when
$1+\alpha\beta$ is sufficiently negative. In this case,
competition between phase instability and forcing leads to the
formation of stripes or hexagonal patterns, with their associated
topological defects. If the forcing is decreased, these structures
break down through spatio-temporal intermittency \cite{kjartan}.
On the other hand, in the phase stable regime, frequency locked
solutions may undergo a variety of bifurcations when forcing is
decreased, leading to oscillation, quasi-periodicity or chaos
\cite{gambaudo}.

The equivalence between the different frequency locked states
makes possible the formation of stable inhomogeneous structures.
These structures are composed of domains of the locked states
separated by abrupt interfaces. Nonrelaxational dynamics may
induce interface motion and, in particular, the formation of
$n$-armed spirals, each arm corresponding to a different frequency
locked solution.

These phenomena were studied in great detail by Coullet and
Emilsson, for $n=1$ and $n=2$, in one- and two-dimensional systems
\cite{kjartan}. The case $n=4$ has been considered theoretically
in \cite{Elphick}. Experimentally, resonant phase patterns
associated with frequency locking have been described for a
periodic forcing of the Belousov-Zhabotinsky chemical reaction for
$n=2$ \cite{Swinney00} and $n=4$ \cite{Swinney01}. For $n=3$, the
existence of three armed rotating spirals in two-dimensional
systems is only briefly mentioned in \cite{kjartan}. Experimental
observations of frequency locking for $n=3$ are also briefly
discussed in \cite{Swinney97,Swinney99}. In these experiments
patterns with three-phase domains shifted by $2\pi/3$ are observed
\cite{Lin01}. It is the aim of this paper to study the case of
resonant forcing for $n=3$ in more detail, and specially the
transition from phase spirals to amplitude spirals. The interest
of this study is threefold. First, it confirms the robustness of
the Ginzburg-Landau dynamics, which is recovered at low forcing,
with all its complexity and its particular sensitivity to kinetic
coefficients. Second, it presents original dynamical behavior in
the excitable regime. This behavior presents interesting analogies
with Rayleigh-B\'enard convection in a rotating cell, described by
three-mode dynamical models \cite{BH,Tu,Gallego}. Third, it could
be a useful framework for the interpretation of detailed
experiments on the 3:1 resonant forcing of a chemical system
\cite{Lin01}.

The paper is organized as follows. The dynamical model and its uniform
asymptotic solutions are presented in section \ref{sec:uniform}. Section
\ref{sec:phaseapprox} is devoted to the description of the dynamics in terms of
phase equations. The properties and possible development of front and spiral
solutions are discussed in section \ref{sec:fronts_spirals}. Numerical results,
for one- and two-dimensional systems, are presented in section
\ref{sec:numerical} and conclusions are drawn in section
\ref{sec:conclusions}.

\section{Uniform solutions}
\label{sec:uniform}

Consider an extended system undergoing a Hopf bifurcation at zero
wavenumber, and subjected to a periodic temporal modulation of
frequency $\omega_e = 3\omega_0$. Sufficiently close to the
bifurcation, its dynamics may be reduced to the following complex
Ginzburg-Landau equation \cite{daniel}:
\begin{equation}\label{gleq3}
\partial_t A= \mu A + (1+i\alpha ) \nabla^2 A - (1+i\beta )
  A\vert A\vert^2 + \gamma\bar A^2,
\end{equation}
where $\gamma\ge 0$ (the case $\gamma<0$ follows by changing $A\to -A$) is
proportional to the external field intensity. The other parameters are standard
\cite{kjartan,daniel}. We will restrict ourselves to this case of resonant
forcing ($\nu=0$). A slightly off-resonant forcing is known to induce a richer dynamical behavior
\cite{gambaudo} whose characterization for a spatially extended system is beyond
the scope of this paper.

We look now for uniform solutions. By dropping the spatial derivative
terms, the corresponding uniform equations are, in phase and amplitude
variables ($A=R_0(t)e^{i\Phi_0(t)}$):

\begin{eqnarray}
\dot{R}_0 &=&\mu R_0 - R_0^3 + \gamma R_0^2 \cos 3\Phi_0, \nonumber\\
\label{null}
\dot{\Phi}_0 &=&- \beta R_0^2 - \gamma R_0 \sin 3\Phi_0.
\end{eqnarray}
If we look at the stationary solutions (fixed points), eqs. (\ref{null}) give
\begin{equation}
(1+\beta^2)R_0^4 - (2\mu + \gamma^2 ) R_0^2 + \mu^2 = 0,
\end{equation}
from where the amplitudes of the uniform solutions are given by:
\begin{equation}
\label{rmm}
R_{\pm}^2= {\frac{1}{2}(1+\beta^2)} [2\mu  +\gamma^2  \pm
\sqrt  {\gamma^4 + 4\mu\gamma^2 - 4\mu^2\beta^2}].
\end{equation}
Such solutions exist provided that $\gamma > \gamma_c$, with
\begin{equation}
\gamma^2_c= 2\mu (\sqrt{1+\beta^2} - 1 ).
\end{equation}
Note that for $\beta=0$ these solutions exist for any nonvanishing
forcing. We will consider the case of $\beta \neq 0$ for which
these solutions appear at a finite value of the forcing. Once the
amplitude is determined by (\ref{rmm}), the phase can be obtained
from the stationary version of (\ref{null}):
\begin{equation}
\cos 3\Phi_0 = \frac{R_0^2-\mu}{\gamma R_0}, \hspace{2.0cm}
\sin 3\Phi_0 = \frac{-\beta R_0}{\gamma}.
\end{equation}
Each value of $R_0=R_+,R_-$ gives rise to three solutions for the phase
$\Phi_0$ which only differ by a phase shift of $2\pi/3$.
Hence,
for $\gamma >\gamma_c$ the system has six uniform solutions:
$(\Phi_1^u,\Phi_2^u,\Phi_3^u)\equiv(\Phi_1^u,\Phi_1^u+2\pi/3,\Phi_1^u+4\pi/3)$ corresponding to $R_-$ and
$(\Phi_1^e,\Phi_2^e,\Phi_3^e)\equiv(\Phi_1^e,\Phi_1^e+2\pi/3,\Phi_1^e+4\pi/3)$ corresponding to $R_+$.
A linear stability analysis shows
that the $\Phi_i^u$ are
always linearly unstable whereas the $\Phi_i^e$
are stable for $|\beta|<\sqrt 3$. The three $\Phi_i^e$ solutions
are called the frequency locked solutions. These become
oscillatory unstable ($k=0$, $\omega\neq 0$) for
$\vert\beta\vert>\sqrt{3}$ in the range of forcing amplitudes
$\gamma_c\lesssim\gamma_1<\gamma<\gamma_2$ where
\begin{eqnarray}
\gamma_1 & = &  \sqrt{\gamma_c^2+\frac{\mu}{2(3\beta^2-1)}
\left[4\sqrt{1+\beta^2}(1-3\beta^2)+
      7\sqrt{3}\beta^3-\beta^2+3\sqrt{3}\beta-5
\right]}
   \nonumber \\
\gamma_2 & = & \sqrt{\mu(1+\beta^2)/2}.
\end{eqnarray}

\begin{figure}
\centerline{\epsfig{figure=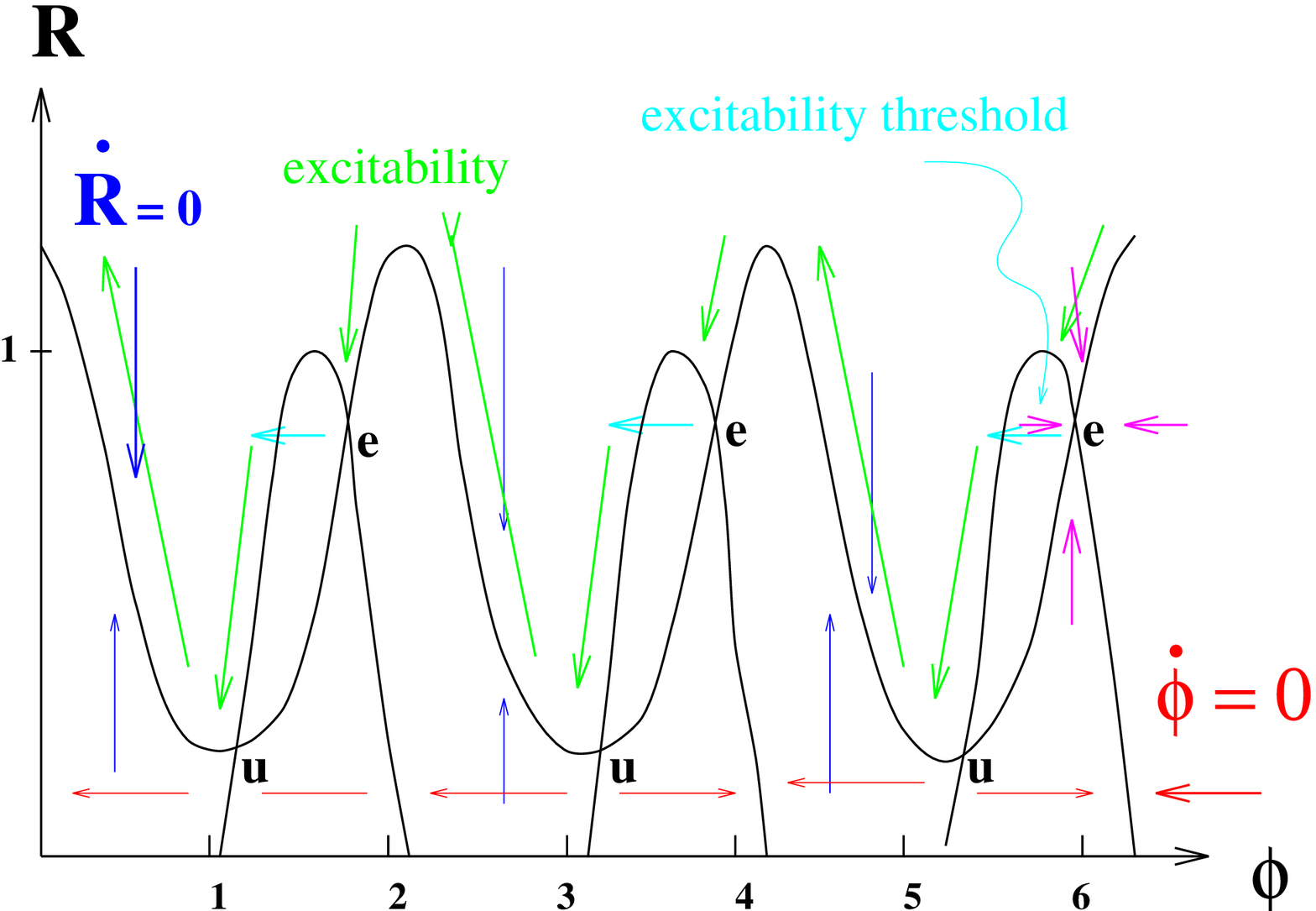,width=10cm}}
\caption{Nullclines and fixed points of the forced Ginzburg-Landau
equation for $\gamma = \beta = 1$, $\mu = 0.25$ in the $R,\phi $
plane. Black arrows indicate the dynamical flow along the
nullcline $\dot R=0$, while grey arrows indicate the dynamical flow
along the nullcline $\dot\Phi=0$ \label{fig:nullA}}
\end{figure}

\begin{figure}
\centerline{\epsfig{figure=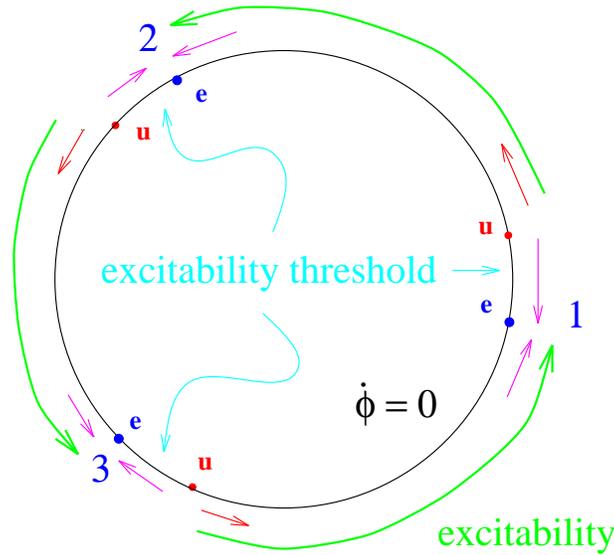,width=8cm}} \caption{Nullcline
$\dot R=0$ in the complex $A$ plane. Along this nullcline the fixed
points $u$ and $e$ are determined by $\dot\Phi=0$. Parameter
values are $\gamma = \beta = 0.01$, $ \mu = 0.25$. Note that for
these parameter values $R_+\approx R_-$. Arrows indicate the
dynamical flow, with long arrows indicating the excitable
excursions associated with an heteroclinic trajectory.
\label{fig:nullB}}
\end{figure}

\begin{figure}[htp]
\centerline{\epsfig{figure=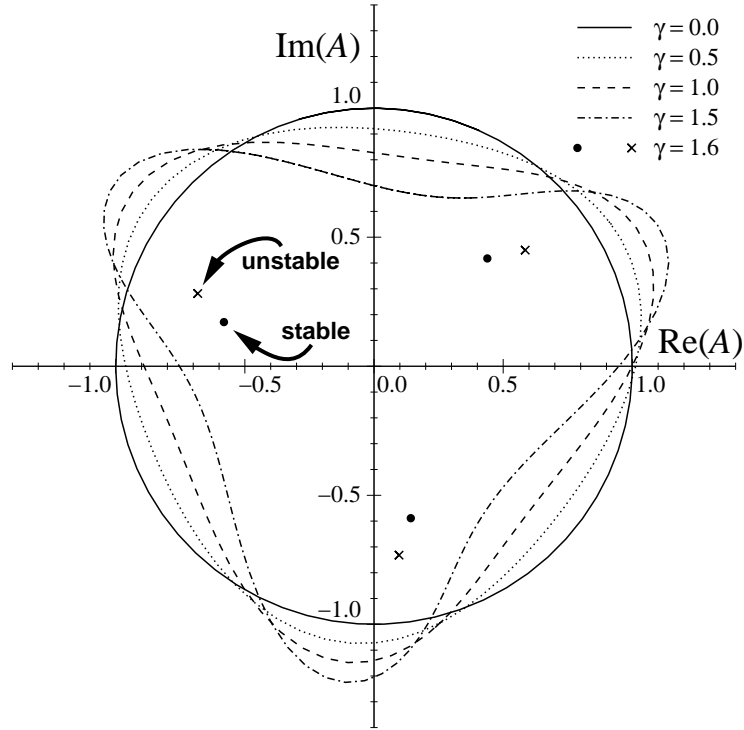,width=10cm}}
\caption{Uniform solutions of eq. (\ref{null}) for several values of the
forcing parameter $\gamma$. System parameters are $\mu=1$, $\beta=-2.0$ (so
$\gamma_c=1.57$). Note that for $\gamma<\gamma_c$ uniform solutions are of the
limit cycle type whereas for $\gamma>\gamma_c$ they become fixed points.
\label{fig:uniform}}
\end{figure}

In the case where the frequency locked solutions are stable, we
can show that the system behaves as an excitable one: let us
construct the nullclines of the dynamical system (\ref{null}),
defined as the curves $\dot R_0=0$ and $\dot\Phi_0=0$, and
represent them in figure \ref{fig:nullA}, for $\gamma = \beta =
1$, $\mu = 0.25$, or in figure \ref{fig:nullB}, for $\gamma =
\beta = 0.01$, $\mu = 0.25$. In both figures, it is easy to see
that the $R_-$ states (labeled $u$) are unstable, while the $R_+$
states (labeled $e$) are stable for small perturbations. However,
for perturbations larger than a well defined threshold, the latter
are unstable and the system makes an excursion in the phase space,
before reaching another, equivalent, steady state. It is a form of
excitability. For $\gamma<\gamma_c$ there are no fixed points of
(\ref{null}) and asymptotic solutions correspond to temporal
oscillations of the limit cycle type. For $\gamma=0$ the limit
cycle is a circle that becomes deformed for $0<\gamma<\gamma_c$
(see figure \ref{fig:uniform}). On increasing $\gamma$, the period
of the oscillations increases and diverges for $\gamma \to
\gamma_c$. The stability of these oscillatory solutions can be
better analyzed in the framework of phase equations that we will
develop in the next section.

The transition between locked and oscillatory states is similar to
the Andronov-van der Pol bifurcation\cite{avk66}, that appears in
several types of excitable systems\cite{mcw94}. In fact, on
decreasing the amplitude of the forcing, stable and unstable
locked states merge via inverse saddle-node bifurcations which
give rise to the birth of limit cycle oscillations. Note that we
have here three pairs of fixed points merging, instead of just one
pair in classical cases. When $\gamma$ approaches $\gamma_c$ from
below, the period of the oscillations diverges as
$(\gamma_c-\gamma)^{1/2}$ showing some kind of critical slowing
down. On the other hand, when $\gamma>\gamma_c$, small
perturbations around the stable states decay, while sufficiently
large perturbations put the system on an heteroclinic trajectory
that goes from an stable fixed point to a different, but
equivalent fixed stable point through the excitability threshold.
The intensity of the perturbations leading to an heteroclinic
trajectory decreases when $\gamma$ approaches $\gamma_c$, and this
phenomenon reflects a type of excitable regime.

All these features can be analytically shown in the limiting case $\beta,\gamma
\ll \mu$. In this limit, and taking into account that $\gamma_c \simeq \beta
\sqrt \mu$, the adiabatic elimination of the amplitude
in (\ref{null}) leads to the phase equation

\begin{equation}
\label{eq10}
\dot\Phi_0 = - \sqrt{\mu}(\gamma_c+\gamma \sin 3\Phi_0),
\end{equation}
from where the excitable stable steady states are given by $\sin
3\Phi_i^e = - \gamma_c/\gamma$ and $\cos 3\Phi_i^e >0$, and
the three unstable steady states satisfy $\sin 3\Phi_i^u =
-\gamma_c/\gamma$ and $\cos 3\Phi_i^u <0$, $i=1,2,3$. In
this case the threshold of the perturbation leading to
excitability is thus given by:

\begin{equation}
\Delta\Phi=\lvert\Phi_i^e -\Phi_i^u\rvert\simeq
\frac{\pi}{3}-\frac{2}{3}\frac{\gamma_c}{\gamma}.
\end{equation}

Equation (\ref{eq10}) describes a relaxational motion of a fictitious test particle in a potential
\begin{equation}
\dot \Phi_0= -\gamma_c\sqrt{\mu}\frac{\partial V(\Phi_0)}{\partial \Phi_0}
\end{equation}
where the potential is
\begin{equation}
V(\Phi_0)=\Phi_0-\frac{\gamma}{3\gamma_c}\cos{3\Phi_0}
\label{potential}
\end{equation}
This potential picture gives a framework to describe the
bifurcation at $\gamma=\gamma_c$ in the language of phase
transitions. In the case $\gamma<\gamma_c$ the potential is
unbounded and has no local minima. As a consequence, the motion is
such that the phase $\Phi_0(t)$ decreases monotonically and the test particle
 does not stop in any selected value of the angle.
When considered modulus $2\pi$, the trajectory is a periodic one $\Phi_0(t+T)=\Phi_0(t)+2\pi$ with a period
$T=(2\pi/\sqrt{\mu})(\gamma_c^2-\gamma^2)^{-1/2}$ which diverges
as $T\sim (-\epsilon)^{-1/2}$ when $\gamma\to\gamma_c^-$ with
$\epsilon=\gamma/\gamma_c-1$. As in mechanical problems, it is possible to
define a probability density function, $P(\Phi_0)$, for the angle variable
as proportional to the time the particle spends in the neighborhood of any
value of the angle (in other words, the probability is inversely proportional
to $\dot \Phi_0$). This probability density function $P(\Phi_0$) develops
three peaks which sharpen as $\gamma$ increases, indicating three preferred
values for the angle. This shows that the characteristic time around each
 minimum increases for $\gamma\to\gamma_c$ and actually it diverges as
 $(-\epsilon)^{-1/2}$. However, for all values of $\gamma<\gamma_c$,
 the dynamics is such that none of the preferred values for the angle is
 actually selected since the system moves continuously from one to another.
 In that sense, we can say that there is a dynamical restoring of the angular symmetry.

The situation is quite different for $\gamma>\gamma_c$. In this case, the
potential $V(\Phi_0)$ develops local minima which, at lower order in $\epsilon$
are $\Phi_1^e=\pi/2+(\sqrt{2}/3)\epsilon^{1/2}$, $\Phi_2^e=\Phi_1^e+2\pi/3$,
$\Phi_3^e=\Phi_1^e+4\pi/3$. These minima correspond to the three frequency
locked solutions. The relaxational motion in the potential is such that for
each trajectory, and depending on the initial condition, the test particle
selects asymptotically one of the local minima and the angular symmetry is now
broken. The probability density function for that particular trajectory becomes
a delta function centered around the selected value of the angle. For an
ensemble of different initial conditions, the probability density function is a
sum of three deltas, $P(\Phi_0)=\frac{1}{3}\sum_{i=1}\delta(\Phi_0-\Phi^e_i)$.

As usual, the transition between the symmetric and the
symmetry-broken phases can be characterized by an order parameter.
For an angular variable, the order parameter should be some
periodic function, such as the mean value of the sine, $\langle
\sin(\Phi_0)\rangle$ (other periodic functions give similar
results). For $\gamma < \gamma_c$ it is $\langle
\sin(\Phi_0)\rangle=0$ whereas for $\gamma > \gamma_c$ that
average sets into one of three possible values corresponding to
the selected phase. Since its value in any phase is different from
zero at $\gamma=\gamma_c$, the order parameter changes
discontinuously at the transition point and, following the
standard notation, the transition can be classified as a
first-order one. In this mean-field description, and according to
the discussion above, the transition implies a critical slowing
down with a characteristic time diverging with a exponent $1/2$
both below and above the transition point. Equivalently, one can
characterize the transition point $\gamma=\gamma_c$ by the fact
that the frequency of the motion tends to zero continuously at
that point and stays equal to zero for $\gamma\ge \gamma_c$. In an
extended system, one could observe coexistence of the three phases
at different locations in space, as described later in figure
\ref{fig:snapsCaseI}. Furthermore, in any finite system, there is
no true phase transition and the sharp behavior predicted by this
simple mean-field analysis gets smeared out and the delta-type
probability density functions for $\gamma>\gamma_c$ have a finite
width. Evidence of this fact is given later from our numerical
simulations (see the lower row of figure \ref{fig:phasePortrait}).

\section{Phase approximation}
\label{sec:phaseapprox}

In this section we present several phase equations each one valid in a
different region of parameters. As mentioned in the previous section,
phase equations can be used to analyze the stability of the uniform
patterns.

\subsection{The oscillatory regime}

In the oscillatory regime ($\gamma<\gamma_c$), the phase dynamics can
be obtained by perturbing the uniform solution $(R_0(t),\Phi_0(t))$, and writing
$R = R_0(t) + \rho (\vec r, t)$, $\Phi = \Phi_0(t +\phi(\vec r,t))$. Following
Hagan \cite{hagan}, the adiabatic elimination of the amplitude perturbations
in the regime $\beta,\gamma\ll\mu$ leads to the following phase dynamics:
\begin{equation}
\partial_t\phi = (1+ \alpha \bar\beta )\nabla^2 \phi +
\kappa (\vec\nabla\phi)^2 + ...,
\end{equation}
where $T$ is the period of the oscillations, and
\begin{equation}
\bar\beta = \frac{\int_0^T dt \, \frac{2\beta R_0 + \gamma \sin 3\Phi_0}{2R_0 -
\gamma\cos \Phi_0}\, \dot\Phi_0^2}{\int_0^T dt\, \dot\Phi_0^2}, \quad
\kappa = \frac{\int_0^T dt\,(\frac{2\beta R_0 + \gamma \sin 3\Phi_0}{2R_0 -
\gamma
\cos \Phi_0}-\alpha )\, \dot\Phi_0^3}{\int_0^T dt\, \dot\Phi_0^2}.
\end{equation}

\noindent For $\gamma \to 0$, one recovers the usual Burgers equation

\begin{equation}
\partial_t\bar\phi = (1+ \alpha \beta )\nabla^2 \bar\phi +(\alpha - \beta
)(\vec\nabla\bar\phi )^2 + ...
\end{equation}
with $\bar\phi = \beta\mu\phi$.

Hence, in the regime where $1+\alpha\bar\beta >0$, stable (phase) spiral waves
may be expected, with wavenumber proportional to $\kappa$, and thus depending
on the characteristics of the oscillations \cite{hagan,Yamada}. In this regime,
the qualitative behavior and interaction between these topological defects
should thus be almost insensitive to the forcing \cite{rica,kramer,pismen}.
Furthermore, in the regime where $1+\alpha\bar\beta < 0$, defect mediated
turbulence should also be expected \cite{coulletgilega}. In the oscillatory
regime, the system presents thus qualitatively the same complexity and the same
spatio-temporal behaviors than self-oscillating systems. Only quantitative
aspects are affected by the forcing.

\subsection{The excitable regime}

In the excitable regime $\gamma>\gamma_c$, and the phase dynamics
can be obtained in the limit $\beta,\gamma \ll \mu$, $\beta\ll 1$
by eliminating adiabatically the amplitude of the field. Taking
into account that, in this regime,  $R^2\simeq\mu$ and
$\gamma_c\simeq |\beta|\sqrt\mu$, we are left with the following
phase equation:

\begin{equation}\label{phasedyn}
\partial _t\Phi = -\sqrt{\mu} (\gamma_c + \gamma  \sin 3\Phi )+
(1+ \alpha \beta )\nabla^2 \Phi -(\alpha - \beta )(\nabla\Phi  )^2
 + \frac{\alpha^2(1+\beta^2)}{2\mu}\nabla^4\Phi
\end{equation}
Besides the homogeneous solutions discussed in the previous section, this
equation admits front solutions connecting stable states asymptotically at
$x=\pm \infty$. In the case $\alpha=\beta= 0$ the phase equation is
relaxational and the fronts connect two states with the same value of the
potential and are, therefore, stationary. In the case $\alpha=\beta\ne 0$, the
phase equation is still relaxational but now the steady states have different
value of the potential and the front moves. Moreover, when $\alpha\ne \beta$
there is a purely nonpotential induced front motion. Equation (\ref{phasedyn})
will be used in the next section as the starting point to compute the velocity
of the front solution.

In order to study pattern forming instabilities, we can use (\ref{phasedyn}) in
the limit of small $\Phi$. Expanding the $\sin 3\Phi$ up to linear order in
$\Phi$, we are led to a damped Kuramoto-Sivashinsky phase
equation~\cite{kjartan}. It follows that frequency locked solutions are stable
for $1+\alpha\beta >0$. If $1+\alpha\beta <0$, a pattern forming instability of
the locked states would occur for~\cite{kjartan}\footnote{This corrects the
missprint of reference \cite{kjartan} in $\epsilon$ and $k_0$ after eq. (28).}

\begin{equation}
\mu > 36\gamma^2\frac{\alpha^4(1+\beta^2)^3}{(1+\alpha\beta )^4}.
\end{equation}
Since, in this regime, $\gamma>\gamma_c$, a necessary condition for this
instability is thus

\begin{equation}
(1+\alpha\beta )^4 > 72(\sqrt{1+\beta^2} - 1 ) \alpha^4(1+\beta^2)^3,
\end{equation}
and this condition cannot be realized in the $\{1+\alpha\beta<0\}$ domain.
Therefore, the frequency locked solutions are stable so that pattern forming
instabilities are ruled out within the phase approximation. It is also possible
to prove that the locked solutions are always stable in the limit of large
forcings~\cite{kjartan}.

\section{Fronts and Spirals}
\label{sec:fronts_spirals}

For $\gamma > \gamma_c$, the forced Ginzburg-Landau equation
possesses three equivalent excitable steady states. The excitability mechanism
described in the previous section provides a natural way of building fronts
between these steady states. Despite the equivalence of the fixed points, such
fronts are expected to move, as a result of the nonpotential character of the
dynamics.

\subsection{One-dimensional systems}

\begin{figure}
\centerline{\epsfig{figure=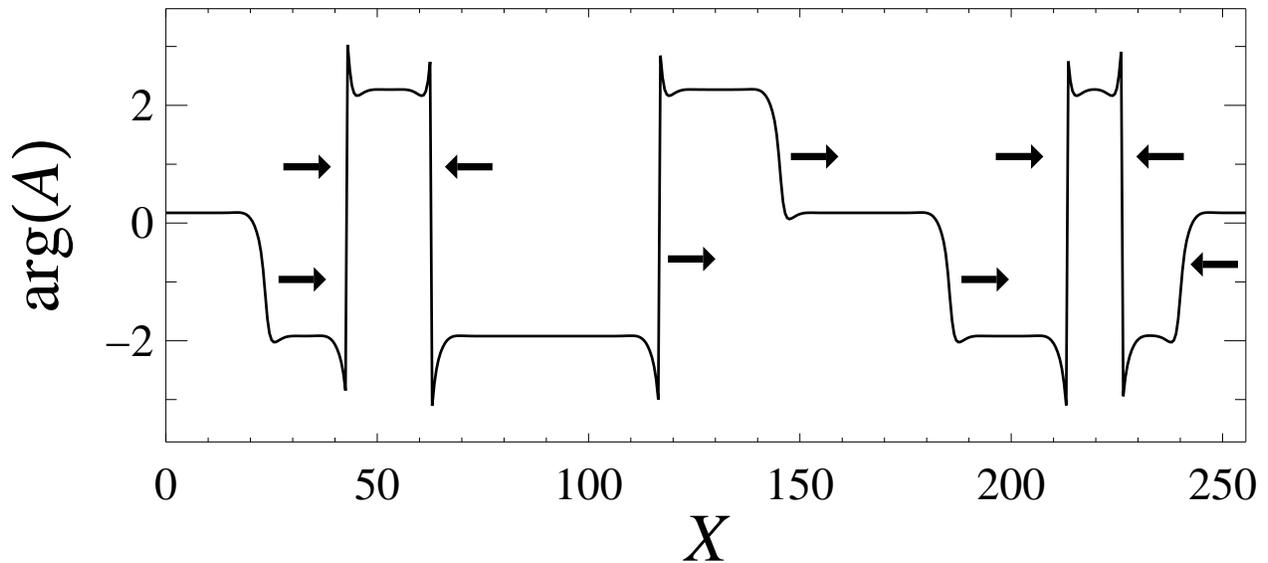,width=\textwidth}}
\caption{Plot of the phase field in the excitable regime in 1d. Note the
existence of three homogeneous phase states. The arrows indicate the direction
of motion of the several fronts. Parameter values are $\mu=1$, $\alpha=2$,
$\beta=-0.2$, $\gamma=0.5$. 
\label{fig:phase1d}}
\end{figure}

Consider a front solution of eq. (\ref{phasedyn}), e.g. $\Phi_{12}(x - vt)$,
joining the states 1 (at $x \to -\infty $) and 2 (at $x \to + \infty$), such
that $\Phi_2^e > \Phi_1^e$. Its velocity may be computed along the standard
procedures, and is such that (at leading order in perturbation)

\begin{equation}\label{vel}
 v =  \frac{ 2\pi\gamma_c\sqrt{\mu}
+3(\alpha - \beta )\int^{+\infty}_{-\infty} (\partial_x \Phi_{12})^3}{
3\int^{+\infty}_{-\infty} (\partial_x \Phi_{12})^2}.
\end{equation}
Hence, for $\alpha >\beta$, the fronts $\Phi_{12}$, $\Phi_{23}$ and $\Phi_{31}$
move to the right, while the fronts $\Phi_{21}$, $\Phi_{13}$ and $\Phi_{32}$
move to the left. Hence, any domain of one steady state, embedded in a domain
of another one, either expands or shrinks, leaving the system in one steady
state (domains of 2 embedded into 1, 3 into 2 and 1 into 3 shrink  while
domains of 1 into 2, 2 into 3 and 3 into 1 expand). However, a succession (from
left to right) of domains with states in the order 1, 2, 3, 1, etc. moves as a
whole to the right. When it is in the order 1, 3, 2, 1, etc., it moves as
a whole to the left (see figure \ref{fig:phase1d}).

\subsection{Two-dimensional systems}

\begin{figure}
\centerline{\epsfig{figure=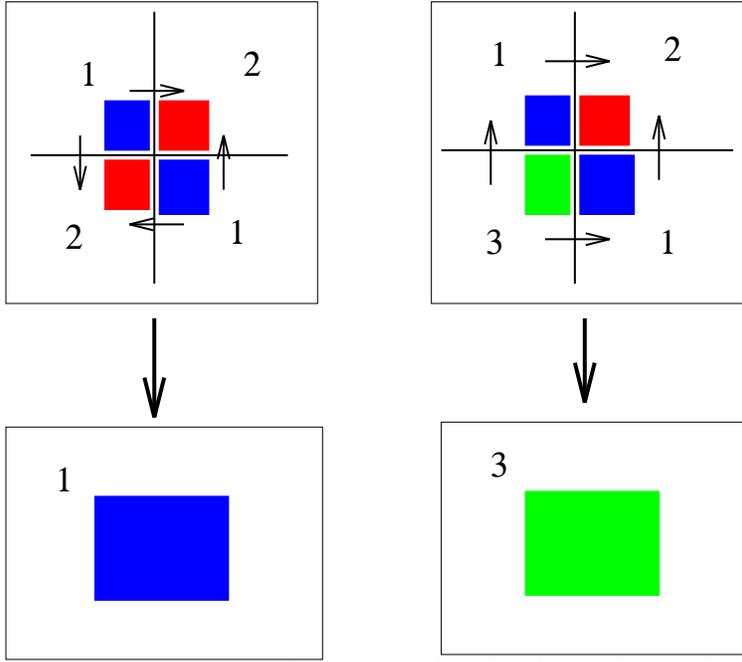,width=10cm}}
\caption{Examples of motion of pairs of inclined fronts separating domains
with equivalent steady states of the forced Ginzburg-Landau equation for
$\gamma >\gamma_c$.
\label{fig:2dfronts}}
\end{figure}

In two-dimensional systems, straight linear fronts have the same behavior as
in one-dimensional systems. Furthermore, sets of two inclined fronts
separating domains with  different steady states also move away or
annihilate, leaving the system in one steady state only (see figure
\ref{fig:2dfronts}).

New phenomena may arise when the three steady states coexist in the system. In
this case, three fronts, which separate the respective domains, coalesce in one
point (a vertex). The three fronts are expected to rotate around this
point. The result is a rotating spiral whose angular velocity increases with
the forcing amplitude. Spirals corresponding to sequences of states in the
order $1\to 2 \to 3$ or  $1\to 3 \to 2$ around the center have opposite senses
of rotation. Isolated vertices remain immobile, but non-isolated ones have a
dynamical evolution induced by mutual interactions, which may even lead to the
annihilation of counter-rotating spirals. This dynamical behavior is
illustrated by the results of the numerical analysis presented in the next
section.

The situation is similar to that observed in system with competing fields.
For example, in the context of fluid dynamics, a three mode model has been
proposed to study Rayleigh-B\'enard convection in a rotating
cell~\cite{BH,Tu}. The fields represent the amplitudes of three set of
convection rolls oriented $60^{\circ}$ at each other. In the two-dimensional
system, vertices may form when the three different types of roll domains meet
at one point (notice that this is not possible in one dimensional
systems~\cite{Gallego}). Then, the nonpotential dynamics induces the
rotation of the interfaces around the vertices preventing the system from
coarsening. At long time scales, the vertices diffuse throughout the system.

\section{Numerical results}
\label{sec:numerical}

In this section, we present numerical results in two spatial
dimensions which illustrate the various dynamical regimes
described in the preceding sections. We have solved numerically
the forced CGLE in two spatial dimensions by using a
pseudospectral method with periodic boundary conditions. We
discretize the system in a square mesh of $256\times 256$ points.
Cases within and beyond the validity of the phase approximation
and with $1+\alpha\beta> 0$ and $1+\alpha\beta<0$ are considered.
In all cases the parameter $\mu$ is taken fixed $\mu=1$. In table
\ref{numericalcases} we summarize the parameters chosen for each
of the cases studied.

\begin{figure}
\centerline{\epsfig{figure=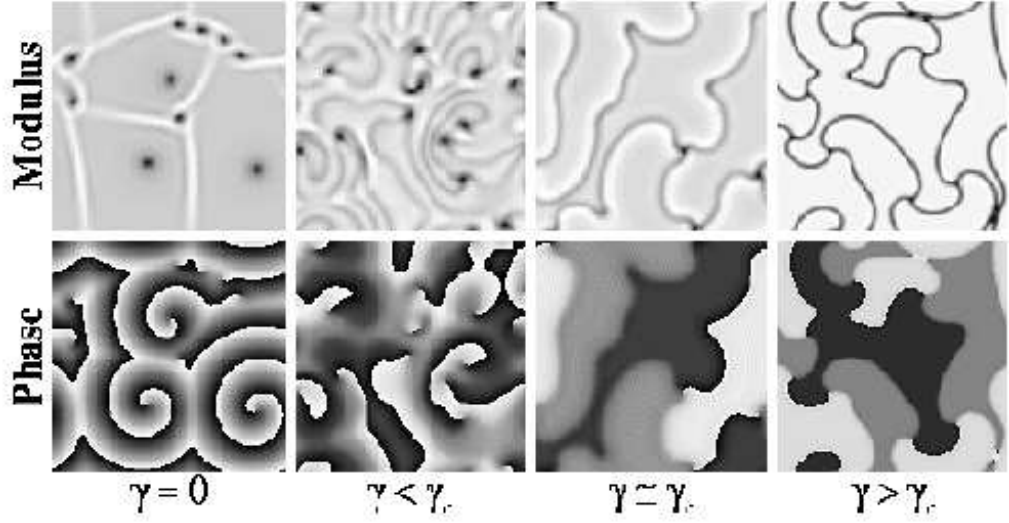,width=0.9\textwidth}}
\caption{Modulus and phase of the complex field $A$ in the cases
$\gamma=0$ (no forcing), $\gamma<\gamma_c$ (oscillatory),
$\gamma\simeq\gamma_c$ and $\gamma>\gamma_c$ (excitable).
Parameter values are $\mu=1$, $\alpha=2$, $\beta=-0.2$ (so that
$\gamma_c\simeq 0.2$), and $\gamma=0.1\ (0.25)$ for the
oscillatory (excitable) case. \label{fig:snapsCaseI}}
\end{figure}

We start our discussion considering cases in which the phase
approximation is valid ($\beta,\gamma\ll \mu$). We first consider
a case below the Benjamin-Fair (BF) line ($1+\alpha\beta>0$)
choosing parameter values \{$\alpha=2$, $\beta=-0.2$;
$\gamma_c=0.2$\}. This case corresponds to the \emph{frozen
states} regime in the phase diagram of the CGLE with no forcing
\cite{Chate}. In figure \ref{fig:snapsCaseI} we show the modulus
and the phase of the complex field $A$ for values of $\gamma$
corresponding to no forcing, oscillatory, $\gamma\simeq\gamma_c$
and excitable regimes. As expected~\cite{Chate}, spiral defects
surrounded by shock lines occur when there is no forcing. When the
strength of the forcing $\gamma$ is increased, but still being
below the critical value $\gamma_c$ (oscillatory regime), the
phase dynamics does not change significantly. However, amplitude
spirals appear in the modulus of the field. The splitting of the
phase into three locked states is observed approximately at the
predicted theoretical value of the forcing $\gamma_c$. For a value
of the forcing parameter slightly greater than $\gamma_c$, we
observe annihilation of vertices until an homogeneous state is
reached. For values of $\gamma$ close to $\gamma_c$ the motion of
the walls can be very slow with patterns that might look
stationary in short time scales of observation. For larger values
of the forcing parameter, the nonrelaxational dynamics is able to
stop vertex annihilation and therefore the coarsening process. The
system remains in a self-sustained dynamical state dominated by
three-armed rotating spirals.

\begin{figure}
\centerline{\epsfig{figure=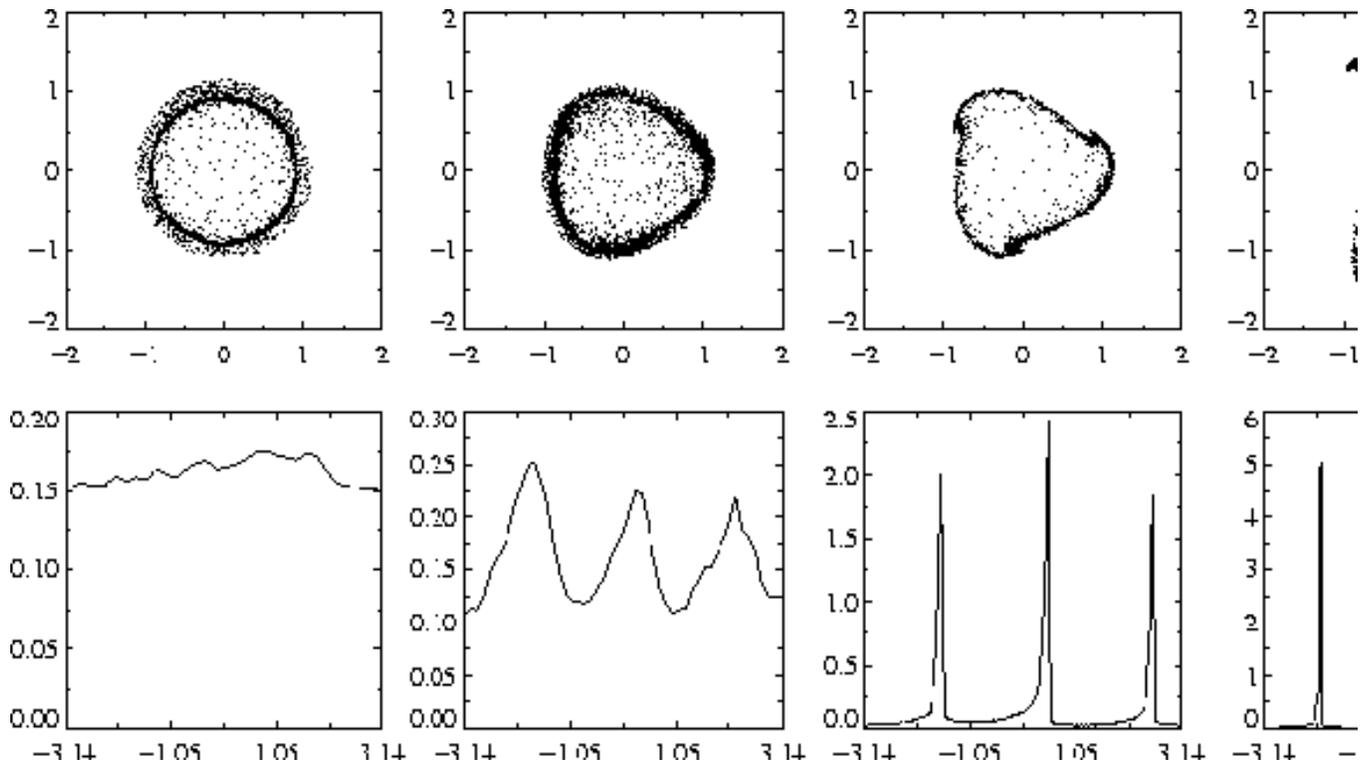,width=\textwidth}}
\vspace{3.0truecm}
\caption{Phase portrait (upper row) in the complex plane of the field $A$ and
corresponding histograms (lower row) of the phase field. Plots have been made
from the snapshots of figure \ref{fig:snapsCaseI}.
\label{fig:phasePortrait}}
\end{figure}

\begin{figure}
\centerline{\epsfig{figure=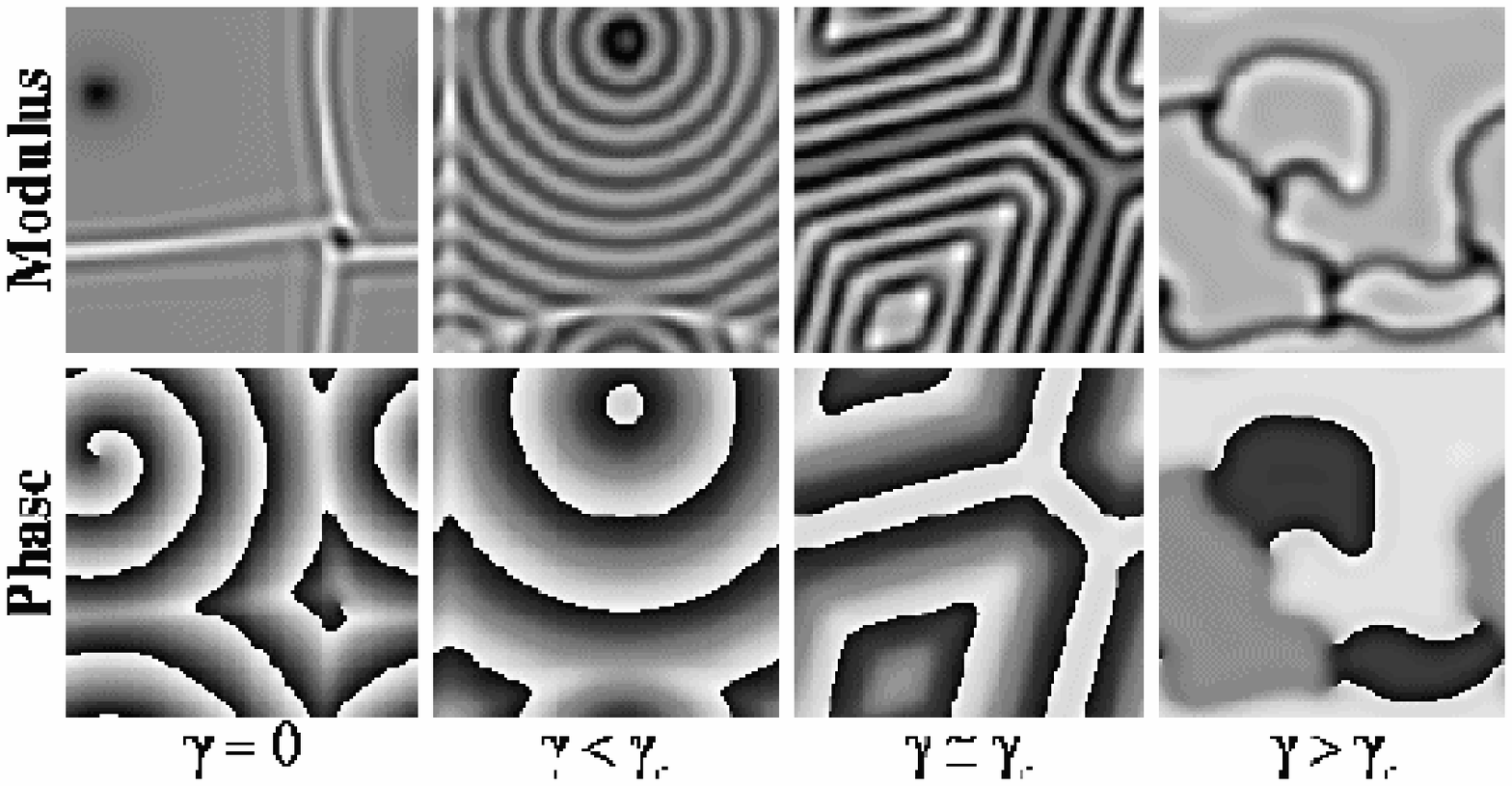,width=0.9\textwidth}}
\caption{Same as in figure \ref{fig:snapsCaseI}. Parameter values are
$\mu=1$, $\alpha=5.5$, $\beta=-0.2$ ($\gamma_c\simeq 0.2$), and
$\gamma=0.1\ (0.25)$ for the oscillatory (excitable) case.
\label{fig:snapsCaseVI}}
\end{figure}

Motivated by the way in which experimental data for 2:1 or 4:1
resonant forcing is presented in \cite{Swinney00,Swinney01}, we
show in figure \ref{fig:phasePortrait} a phase portrait in the
complex plane of $A$, obtained from the snapshots shown in figure
\ref{fig:snapsCaseI}. We also include the corresponding histograms
for the values of the phase of $A$. This representation of data
clearly displays the transition from the oscillatory to the
excitable regime. The phase portrait gives a demonstration, for
the spatially extended system, of the deformation of the limit
cycle described in figure \ref{fig:uniform}. Furthermore, it
enlightens the transition from one-armed spirals, which are
generic defects of unforced oscillations, to three armed spirals,
which are generic defects of the locked states. Effectively, at
$\gamma=0$, one observes typical phase spirals associated with the
monotonic phase variations. In this case, there are no fronts
separating different states of the system. When $\gamma$
increases, the phase variable spends an increasing amount of time
around the precursors of the locked states, as reflected in the
histograms of Fig. \ref{fig:phasePortrait}. This manifests itself
in the appearance of fuzzy three armed patterns around vertices of
the modulus of the field $A$ (see for example
Fig.\ref{fig:snapsCaseVI}). For $\gamma$ slightly larger than
$\gamma_c$ the system has undergone the transition form
oscillatory to locked behavior and one may observe spirals
generated by fronts between different stable steady states of the
system. This transition is in fact the spatial unfolding of the
evolution of the phase portrait displayed in
Fig.\ref{fig:phasePortrait} for increasing $\gamma$, which in turn
corresponds to the mean field description of the transition given
after eq. (\ref{potential}).

Above the BF line ($1+\alpha\beta<0$), we choose $\alpha=5.5$ and
keep the rest of parameters as before. The most noteworthy
difference with the previous case is the existence of asymptotic
frozen states for all values of $\gamma$ within the oscillatory
regime, even close to the critical forcing $\gamma_c$ (see figure
\ref{fig:snapsCaseVI}). Below $\gamma_c$, we observe frozen
targets while close to the transition ($\gamma \approx \gamma_c$)
the frozen patterns hold three locked phase states but without
vertices. The particular geometry of these patterns is due to the
emergence of locked phase states. As expected, large enough values
of the forcing parameter give rise to a time-dependent dynamics
with three-armed spirals rotating around vertices.

\begin{figure}
\centerline{\epsfig{figure=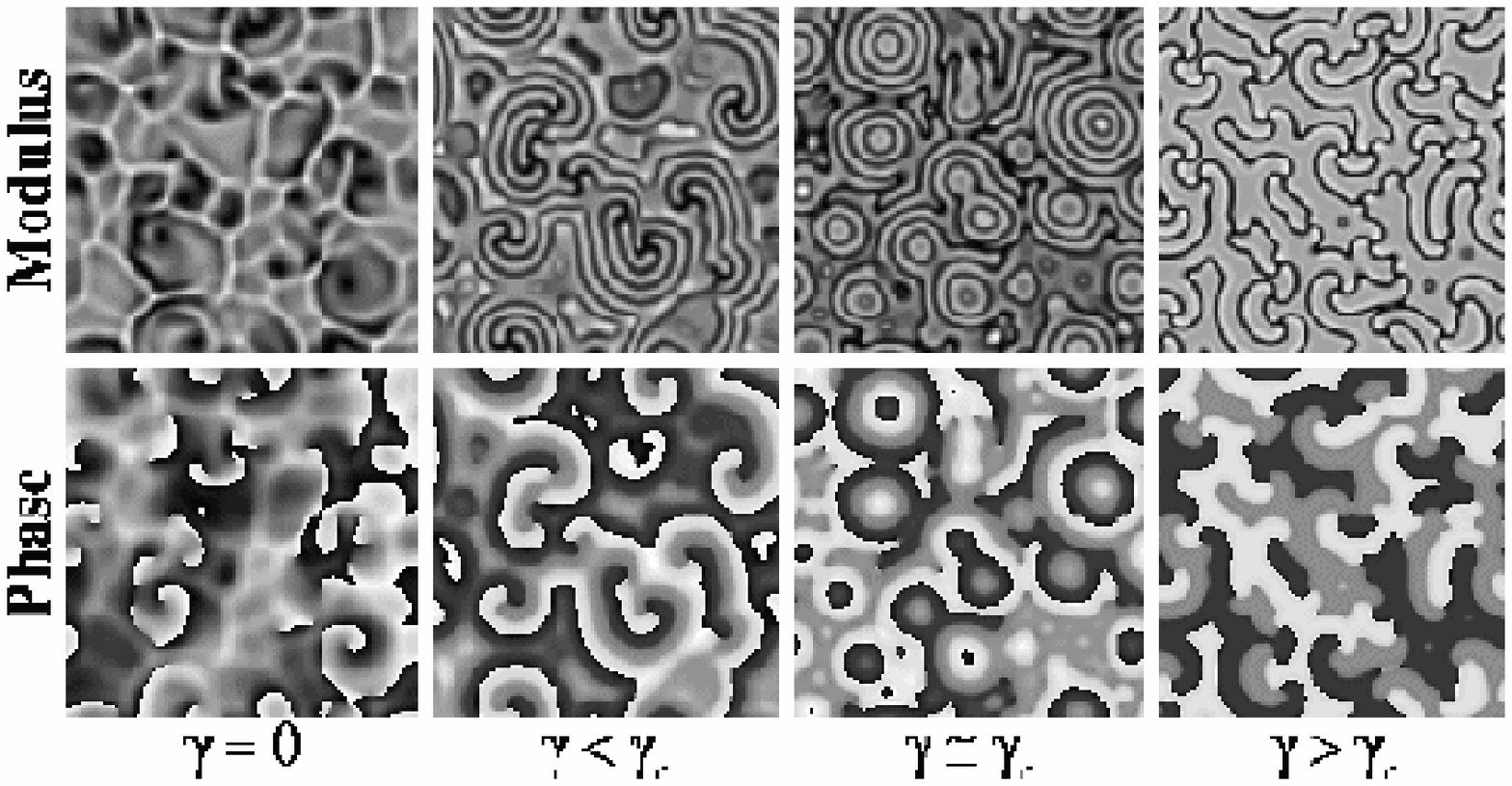,width=0.9\textwidth}}
\caption{Same as in figure \ref{fig:snapsCaseI}. Parameter values are
$\mu=1$, $\alpha=2$, $\beta=-0.76$ ($\gamma_c\simeq 0.72$), and
$\gamma=0.5\ (1.5)$ for the oscillatory (excitable) case.
\label{fig:snapsCaseV}}
\end{figure}

Beyond the validity of the phase approximation, different
phenomena may occur. In particular, pattern forming instabilities
may take place for small and moderate values of the forcing
parameter above its critical value $\gamma_c$. In the case above
the BF line with parameters \{$\alpha=2$, $\beta=-0.76$;
$\gamma_c=0.72$\} (\emph{phase turbulence} regime in the absence
of external forcing), oscillating targets that coexist with
vertices are observed close to the transition (see figure
\ref{fig:snapsCaseV}). Pulses in the modulus of $A$ form at the
center of the targets concentric rings with an amplitude which
decays in space as they propagate away from the center, while the
phase oscillates periodically between $-\pi$ and $\pi$.

\begin{figure}
\centerline{\epsfig{figure=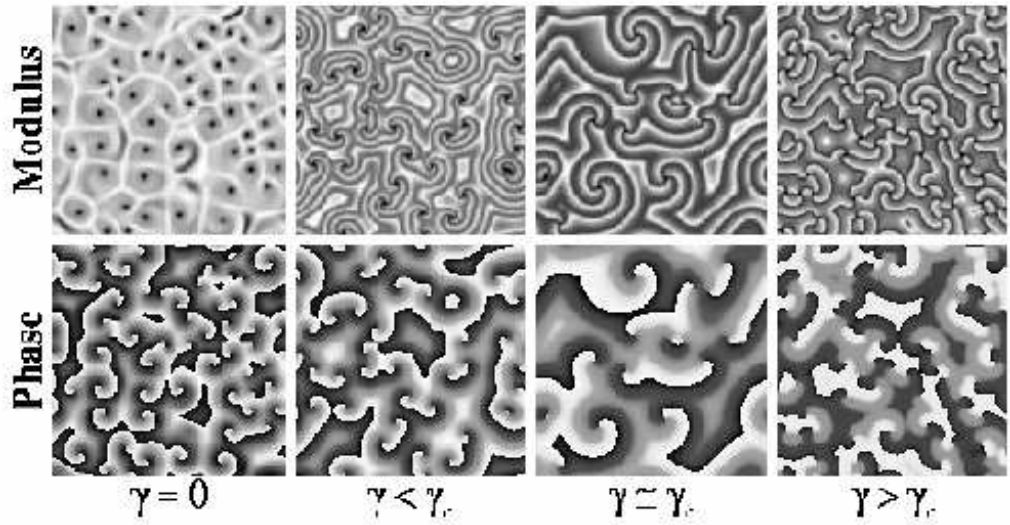,width=0.9\textwidth}}
\caption{Same as in figure \ref{fig:snapsCaseI}. Parameter values are
$\mu=1$, $\alpha=0$, $\beta=-1.8$ ($\gamma_c\simeq 1.45$), and
$\gamma=1\ (1.6)$ for the oscillatory (excitable) case.
\label{fig:snapsCaseIV}}
\end{figure}

\begin{figure}
\centerline{\epsfig{figure=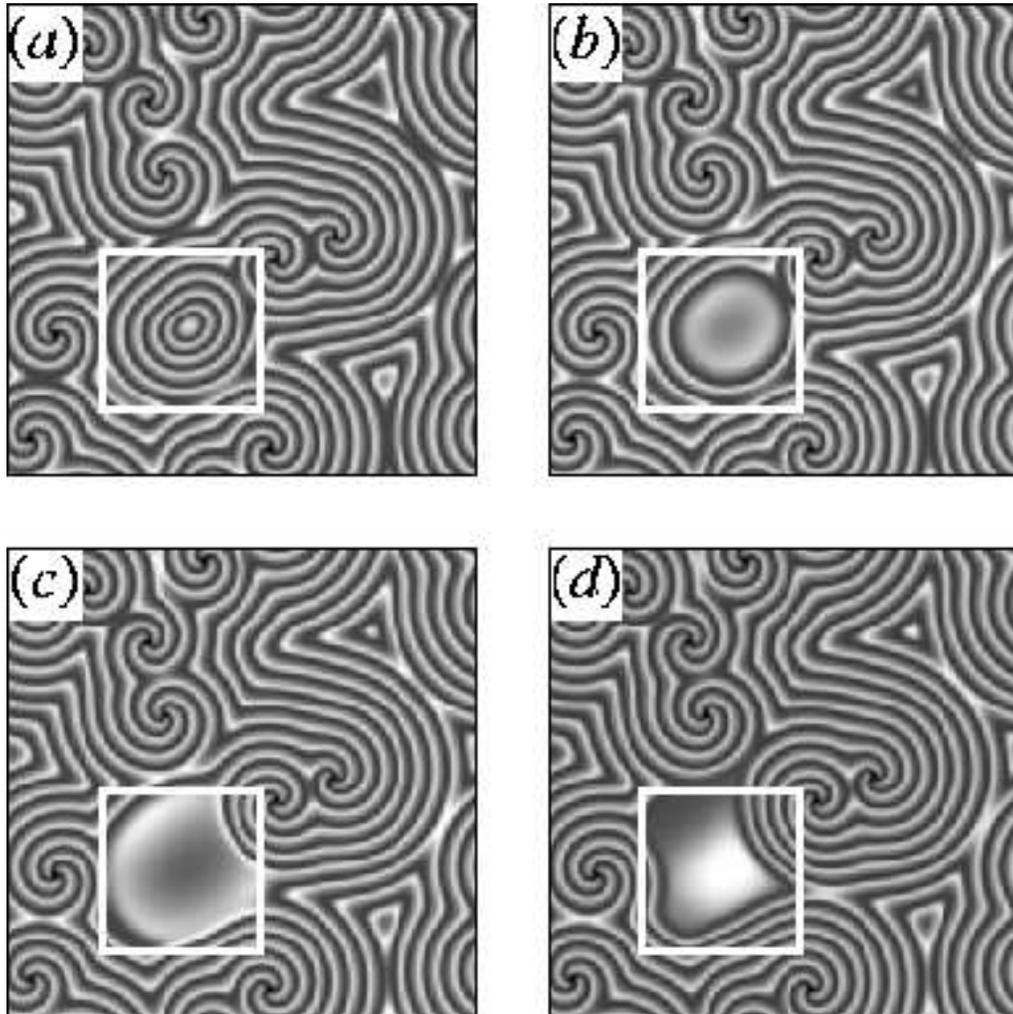,width=0.75\textwidth}}
\caption{Snapshots of the modulus of the field in a regime of
parameters where an oscillatory instability at zero wave number occurs. The
square encloses an oscillating region. Time increases when going from (a)
to (d).
\label{fig:zero}}
\end{figure}

We also considered, beyond the validity of the phase
approximation, a case below the BF line. Our choice of parameters
\{$\alpha=0$, $\beta=-1.8$; $\gamma_c=1.45$\} corresponds to a
\emph{defect turbulence} regime at $\gamma=0$. Since
$1+\alpha\beta>0$, well-developed spirals can be observed (see
figure \ref{fig:snapsCaseIV}). The modulus of the field is
characterized for $\gamma<\gamma_c$ by amplitude spirals that
rotate around defects, whereas the phase shows a behaviour similar
to the one for $\gamma=0$. These three armed amplitude spirals
become well developed for $\gamma>\gamma_c$ corresponding to the
emergence of the three possible homogeneous phase values. On the
other hand, since $\vert\beta\vert>\sqrt{3}$, and according to the
discussion of section \ref{sec:uniform}, there exists a range of
values of the forcing parameter for which the locked solutions
present a oscillatory instability at zero wavenumber. This is seen
in figure \ref{fig:zero}. This instability is observed after the
annihilation of two counter-rotating defects. In the squared
region identified in the figure we observe the development of an
oscillating target corresponding to the homogeneous oscillation.
>From figure \ref{fig:zero}d onwards the oscillating regions
shrinks and disappears under the invasion of neighboring spirals.

It is important to emphasize that for $\gamma\gg\gamma_c$ the
asymptotic state is essentially the same regardless of the
different dynamical regimes that exist for $\gamma=0$ and
different parameter values. The phase is locked to either of the
three values predicted theoretically, with interfaces between
these three locked states rotating around vertices. This rotation,
which is due to the underlying nonrelaxational dynamics, inhibits
the coarsening process  which would take place through vertex
annihilation. The vertices are essentially pinned and the
resulting pattern is, on the average, time periodic at relative
short time scales. When the phase approximation is valid, the
locked phase states are seen to be stable, but excitable spirals
may be absent near the transition for system parameters such that
$1+\alpha\beta<0$. Beyond the validity of the description based on
the phase approximation, instabilities of the homogeneous phase
states may take place giving rise to complex patterns.

\section{Summary}
\label{sec:conclusions}

Temporal forcing of nonlinear self-oscillating extended systems
strongly couple with with the unstable modes associated with a
Hopf instability. Such forcings modify the character of the
bifurcation and the resulting spatio-temporal patterns, as
observed in different resonant regimes of a forced
reaction-diffusion system
\cite{Swinney97,Swinney99,Swinney00,Swinney01}. In this paper we
have studied the particular case of a resonant temporal modulation
at three times ($n=3$) the critical frequency of the Hopf
bifurcation.

For forcing amplitudes $\gamma$ below a critical value $\gamma_c$,
the system is in an oscillatory regime, where the spatio-temporal
behavior strongly depends on the parameters of the associated
Ginzburg-Landau equation. Uniform solutions correspond to temporal
oscillations of the limit cycle type and topological defects
correspond to one-armed phase spirals. When $\gamma$ approaches
$\gamma_c$ from below, the period of the limit cycle diverges as
$(\gamma_c-\gamma)^{1/2}$ showing some type of critical slowing
down. For forcing amplitudes above the critical one, the system is
in a phase locked regime with three equivalent steady states. The
bifurcation occurring at $\gamma=\gamma_c$ can be described, in a
mena field approximation, as a first order phase transition
between a motion in which the temporal average of the phase is
zero ($\gamma<\gamma_c$) to one with a nonzero average phase
($\gamma>\gamma_c$). We have given a phase approximation
description of both regimes for a spatially extended system.

Like in the $n=1$ and $n=2$ cases of strongly resonant forcings, a
form of excitability may also be observed. However, contrary to
the $n=1$ and $n=2$ cases, no pattern forming instability of the
frequency locked states occurs, in this case, for parameter values
for which the phase approximation is valid. Due to the
nonrelaxational character of the dynamics, fronts between
equivalent steady states move. The result is that, when the three
equivalent steady states coexist in the system, three armed
rotating spirals are generated around vertices where the fronts
separating each domain meet. Hence, we predict a transition from
one-armed phase spirals to three-armed excitable amplitude
spirals, which occurs when the forcing amplitude passes through a
critical value $\gamma_c$. We have confirmed and described this
transition by numerical analysis of the corresponding Complex
Ginzburg Landau equation for different parameter values which
correspond to qualitatively different regimes of the phase diagram
of this equation when there is no forcing.

There are a number of analogies of our results for
$\gamma>\gamma_c$ and the spatiotemporal patterns observed in a
model of rotating Rayleigh-B\'enard convection \cite{Gallego}. In
the latter case the domains correspond to sets of paralell
convection rolls with a certain orientation and the vertices to
points at which fronts separating domains of three preferred
orientatons meet. As in the case studied here, the rotation of
interfaces around vertices, due to nonrelaxational dynamics,
produce rotating three armed spirals.

\acknowledgements We thank A. Lin and H. Swinney for sharing with
us information on  their unpublished results on 3:1 resonant
patterns. We acknowledge financial support from DGESIC (Spain) projects
BFM2000-1108 and PB-97-0141-C02-01.

\clearpage

\begin{table}
\newlength{\LL} \settowidth{\LL}{Defect Turbulencex}
\begin{center}
\begin{tabular}{|m{1.2cm}<{\centering}>{\centering}m{1.2cm}>{\centering}m{1.2cm}
>{\centering}m{1.2cm}>{\centering}m{2.75cm}>{\centering}m{1.5cm}
>{\centering}m{\LL}m{1.5cm}<{\centering}|}
\hline $\mu$ & $\alpha$ & $\beta$ & $\gamma_c$ & Phase approx. valid? &
$1+\alpha\beta$ & Regime ($\gamma=0$) & Figure \\ \hline\hline
$1.0$ & $2.0$ & $-0.20$ & $0.20$ & Yes & $>0$ & \emph{Frozen states} &
\ref{fig:snapsCaseI}, \ref{fig:phasePortrait} \\
$1.0$ & $5.5$ & $-0.20$ & $0.20$ & Yes & $<0$ & \emph{Frozen states} &
\ref{fig:snapsCaseVI} \\ \hline
$1.0$ & $2.0$ & $-0.76$ & $0.72$ & No  & $<0$ & \emph{Phase turbulence} &
\ref{fig:snapsCaseV} \\
$1.0$ & $0.0$ & $-1.80$ & $1.45$ & No  & $>0$ & \emph{Defect turbulence} &
\ref{fig:snapsCaseIV}, \ref{fig:zero} \\ \hline
\end{tabular}
\end{center}
\caption{Parameters of the various cases discussed in section
\ref{sec:numerical}.
\label{numericalcases}}
\end{table}


\end{document}